\documentclass[fleqn,10pt]{wlscirep}
\usepackage[utf8]{inputenc}
\usepackage[T1]{fontenc}
\usepackage{multirow}
\title{Prediction of Spallation Induced Transmutation Rates For Long-Lived Fission Products via Proton Accelerator}

\author[1,*]{Grigor Tukharyan}
\author[1, *]{W. Reed Kendrick}
\author[1]{Jiankai Yu}
\author[1]{Areg Danagoulian}
\author[1]{Benoit Forget}
\affil[1]{Department of Nuclear Science and Engineering, Massachusetts Institute of Technology, Cambridge, MA}
\affil[*]{grigt@mit.edu, rkendric@mit.edu}

\begin{abstract}
Long-lived fission products  represent a major challenge in nuclear waste management due to persistent radiotoxicity over very long timescales. This study focuses on six of these fission products: Se-79, Zr-93, Tc-99, Sn-126, I-127, Cs-135. This study investigates the feasibility of spallation-driven transmutation, in which a high energy proton beam strikes a heavy spallation target to generate neutrons that induce transmutation in the fission products surrounding the target. Lead and depleted uranium are identified as the principal spallation target candidates, reflecting contrasting trade offs in neutron yield, secondary reactions, and heat generation. Simulations assess nuclide specific behavior under reactor scale inventories and practical geometric constraints. Results demonstrate that technetium, iodine, and selenium are strong candidates for transmutation using this pathway, while tin shows partial resistance but benefits from thermal flux. By contrast, zirconium is inefficient to transmute, and cesium suffers from low net reduction due to competition with lighter isotopes. Cost effectiveness is highly isotope-dependent: technetium is most favorable, whereas cesium and zirconium remain expensive. These findings highlight the advantages and limitations of spallation driven systems and motivate strategies with optimized target–blanket designs.
\end{abstract}
\begin{document}


\flushbottom
\maketitle

\thispagestyle{empty}

\section{Introduction}

Long-term management of nuclear waste, specifically regarding storage and disposal, continues to be a pressing issue in the United States. At present, most nuclear waste is temporarily contained in spent fuel pools or dry cask storage at nuclear power facilities, yet a well-defined, sustainable, long-term strategy remains absent. This situation perpetuates public apprehension about nuclear fission technology and its implications for environmental and human safety \cite{ref:alvarez}. Central to these concerns are long-lived fission products (LLFPs), a subset of nuclear waste isotopes notable for their particularly long half-lives. These radioactive isotopes primarily result from the fission of uranium or plutonium within nuclear reactors and predominantly emit beta radiation, complicating detection efforts during potential containment breaches or accidents \cite{ref:llfp}. Due to their prolonged radiotoxicity, LLFPs require secure, isolated storage spanning extremely long periods.

Six LLFP nuclides have been identified in a study by the U.S. Department of Energy (DOE) as especially significant, contributing to over 99\% of the residual radiotoxicity of nuclear waste after the recycling of actinides \cite{ref:wigeland}. These nuclides are Technetium-99 (Tc-99), Cesium-135 (Cs-135), Zirconium-93 (Zr-93), Tin-126 (Sn-126), Iodine-129 (I-129), and Selenium-79 (Se-79) \cite{ref:wigeland}. Addressing the problem of LLFPs is increasingly critical as countries look to expand nuclear energy as part of their carbon-neutral energy strategies. Without viable solutions, public perception of nuclear energy risks being continually undermined by concerns about nuclear waste. A notable reduction in long-term radiotoxicity could be achieved through the targeted transmutation of these nuclides into shorter-lived or stable forms, significantly alleviating the burden of long-duration storage.

Many transmutation pathways exist for the selected nuclides, and this work seeks to evaluate neutron transmutation from proton induced spallation. Proton induced spallation presents a promising technique for transmuting LLFPs, because of its capacity to generate intense neutron sources. The neutrons generated through proton-induced spallation processes can be used directly to transmute LLFP isotopes within specialized target assemblies downstream of the primary spallation target. By carefully influencing the neutron energy spectrum through moderators or tailored shielding materials, specific reaction channels can be enhanced, optimizing the transmutation efficiency for different LLFPs \cite{ref:spal2, ref:gary}.

This work focuses on characterizing transmutation rates of LLFPs under neutron flux driven by a 1 GeV proton accelerator. Target material performance and LLFP spectral sensitivity are recorded in order to project estimations of removal rates and economic impact for a proposed target-blanket design.

\section{Background} 

\subsection{Long-Lived Fission Products (LLFPs)}
LLFPs are a category of radioactive materials characterized by exceptionally long half-lives, ranging from hundreds of thousands to millions of years. Nuclides with longer half-lives are generally not a concern as their radioactive activity is too low to contribute much to residual radiotoxicity. These nuclides are produced in nuclear reactors as byproducts of the fission process. The LLFP targets are constructed according to the isotopic compositions found in spent nuclear fuel, where the spent nuclear fuel is assumed to be elementally separated\cite{ref:wigeland}. The mass fractions for every isotope are listed in Supplementary Tables S1-S5, where the mass fraction of each LLFP isotope refers to its relative abundance within the isotopic composition of its parent element. Table \ref{tin} provides the half-lives for each LLFP, as well as the fraction of mass each LLFP makes up in their isotopic composition. 

\begin{table}[ht]
    \caption{\label{tin}Initial isotopic composition mass fraction and half life for each LLFP. The mass fraction of each LLFP isotope refers to its relative abundance within the isotopic composition of its parent element (e.g., the fraction of Se-79 among all selenium isotopes created in the nuclear spent fuel) \cite{ref:wigeland}.}
    \centering
    \begin{tabular}{cccc}\toprule 
    Nuclide&   Half-Life (years)& Mass Fraction
\\ \midrule
    Se-79& 3.27E+05 & $13.88\%$ \\
    Zr-93& 1.61E+06 & $20.26\%$ \\
    Tc-99& 2.11E+05 & $100\%$ \\
    Sn-126& 2.33E+05 & $29.57\%$ \\
    I-129& 1.61E+07 & $72.15\%$ \\
    Cs-135& 1.33E+06 & $38.06\%$ \\
    \bottomrule
    \end{tabular}
\end{table}

 Tc-99 (11.5 g/cm$^3$) is the only isotope of technetium that remains in nuclear reactor spent fuel \cite{ref:transmut}. For Tc-99, the targets are assumed to be metallic, with the soluble fraction converted into metallic technetium. For the other elements, the presence of multiple isotopes in the target material introduces additional complexity and lowers the overall transmutation rates of the LLFPs. 

\subsection{Spallation}

Previous work analyzing transmutation rates of LLFPs using direct proton irradiation showed inefficient performance, highlighting the need of a stronger source of particles \cite{ref:physor}. Proton induced spallation is used as an intense, tunable neutron source for transmutation because it liberates a large number of fast neutrons from a heavy metal target, which can then drive neutron-induced reactions in a downstream absorber. When a high‐energy proton (hundreds of MeV to GeV) penetrates a high-Z nucleus, it first triggers an intranuclear cascade: a femtosecond‐scale shower of energetic nucleons and mesons. After the cascade, the excited residual nucleus emits a larger population of lower-energy neutrons (below 20 MeV) via an evaporation process. By adjusting the incident proton energy and target material, spallation allows precise control over total neutron multiplicity and spectral shape, making it ideally suited to convert long-lived fission products such as Tc-99 via successive neutron captures and threshold (n,xn) channels \cite{ref:spal2}. 

Spallation distinctly differs from fission in both the energy of the emitted particles and the quantity of neutrons generated. Neutrons produced during spallation reactions carry significantly higher energies compared to those from fission, as secondary neutrons frequently maintain energy levels comparable to the initial particle that initiated the reaction. Furthermore, spallation yields a substantially higher number of neutrons compared to fission. For example, a typical fission event releases an average of 2.5 neutrons, whereas a single 800 MeV proton striking a tungsten target can generate around 13 neutrons in a spallation reaction \cite{ref:gary}. 

The initial stage of spallation is the intranuclear cascade, which unfolds on sub-picosecond time scales. The incident proton transfers energy via multiple hard collisions with bound nucleons, generating a 'hadronic shower' of fast neutrons, protons, and pions (above 20 MeV) that escape the nucleus. These high-energy secondaries can themselves induce further spallation events in adjacent nuclei, extending the cascade and increasing the total neutron yield. Approximately 2–15\% of all neutrons emerge directly from this cascade phase, and their forward-peaked angular distribution contributes to anisotropic leakage patterns in realistic target assemblies. The residual nucleus, having lost multiple nucleons, is left in a highly excited state, setting the stage for the slower evaporation phase \cite{ref:spal2}.

Once the cascade subsides, the excited residual nucleus de-excites by evaporating particles, which are mostly neutrons (below 20 MeV), with minor proton and alpha emission channels. The neutron spectrum follows a Maxwellian distribution:
\begin{equation}
N(E) \;\propto\; \sqrt{E}\,\exp\!\Bigl(-\frac{E}{T}\Bigr),
\qquad
T \approx 1.3\;\mathrm{MeV}
\end{equation}
Emission is nearly isotropic in the center-of-mass frame, though lab-frame transformations impart slight forward bias at higher excitation energies. The spectrum is similar to the Watt spectrum created by prompt fission neutrons in a nuclear reactor. For very heavy targets (A \textgreater 
 200), high-energy fission can actually compete with evaporation, releasing additional neutrons and creating fission residues. About 85-98\% of all neutrons are created from this process \cite{ref:spal2}.

Thick-target yield systematics provide a convenient estimate of total neutron production. For non-fissile materials, the relation is as follows:
\begin{equation}
Y(E_p, A) \;=\; 0.1\,\bigl(E_p - 0.120\bigr)\,(A + 20),
\quad E_p \text{ in GeV}
\end{equation}
Where $E_p$ represents the energy of the incoming proton in GeV, and $A$ represents the material's mass number. This captures the linear rise above the 120 MeV threshold as well as the geometric surface effects (A + 20 term) and the energy dependence of the intranuclear cascade \cite{ref:spal2}. However, this relation fails when using a fissile or fissionable material, as the neutron yield almost doubles because of secondary fission reactions. For U-238, the relation is as follows:

\begin{equation}
Y(E_p,A) = 50(E_p - 0.120)
\quad E_p \text{ in GeV}
\end{equation}
The neutron yield for U-238, unlike other heavy nuclei, includes a significant contribution from secondary fission in addition to spallation. Pre-equilibrium neutron emissions also alter the balance between cascade and fission neutrons, so the yield does not scale linearly with the mass number A. Above 1 GeV, both of these relations lose linearity and $E_p$ should be replaced by $E_p^{0.8}$ \cite{ref:spal2}.

Recent research has explored accelerator driven and reactor based systems for the transmutation of LLFPs. Kang et al. demonstrated that coupling a 1 GeV proton beam with a compact subcritical assembly enables efficient Tc-99 transmutation. The neutron spectrum is adjusted with Be or Pb–Bi moderators, achieving high transmutation rates while focusing on trade offs between neutron yield, target activation, and heat load \cite{ref:Kang}. Chiba et al. demonstrated that a YD$_2$ moderated fast reactor can effectively achieve net LLFP reduction in six major isotopes (Se-79, Zr-93, Tc-99, Pd-107, I-129, Cs-135) without isotope separation, showing that moderation in peripheral regions increases neutron capture probabilities and enhances transmutation efficiency \cite{ref:Chiba}. Sun et al. proposed a photonuclear accelerator driven system combining Pb–Bi and Be layers for simultaneous transmutation of multiple LLFPs, finding that neutron spectra can be tuned to balance capture efficiency across isotopes, though at the cost of extremely high beam intensities \cite{ref:anes}. Complementary work by Jin et al. examined proton, deuteron, and $\alpha$ induced spallation on thick LLFP targets, identifying energy windows (50–200 MeV) and projectile choices that maximize reaction yield while minimizing secondary product buildup, underscoring the importance of optimizing target composition and geometry for realistic implementation \cite{ref:Jin}.

Although these studies demonstrate the feasibility of LLFP transmutation using accelerator driven or hybrid systems, most have been limited to geometries that do not capture the coupled effects of neutron spectrum, target composition, and spatial isotope arrangement. The present work expands on these efforts by analyzing the LLFPs within a unified spallation-driven framework. Using Monte Carlo simulations of Pb and depleted uranium spallation targets, this study analyzes the relationships between neutron yield, spectral characteristics, and isotope positioning to determine their combined impact on transmutation efficiency.

\section{Methods}

\subsection{PHITS}
PHITS (Particle and Heavy Ion Transport System), a Monte Carlo particle transport code, is a key tool utilized extensively in this study \cite{ref:phits}. This advanced simulation software is specifically designed to model the transport of radiation across various materials with high precision. PHITS excels at capturing the complex interactions between particles, heavy ions, and matter, including detailed representations of nuclear reactions and electromagnetic processes. In this paper, PHITS is employed to analyze the generated neutron flux spectra and to find the proton penetration depth of proton beams in a target. Every PHITS run initializes and transports 100,000 protons with the following beam characteristics: 1 GeV energy, 30 mA current, in a 1 cm radius. The simulation is run in continuous energy mode and tallies a multigroup flux with the CCFE-709 group structure. The PHITS analysis in this work uses INCL4.6 \cite{ref:INCL} and GEM \cite{ref:GEM} for high energy proton reaction data, as well as JENDL-4.0 for neutron reaction data \cite{ref:JENDL}.

Although substantial experimental data exist for proton induced spallation reactions in materials such as lead and tungsten \cite{ref:spal2}, allowing reliable neutron yield benchmarking, the neutron induced cross sections of the LLFPs present a significant source of uncertainty. Experimental validation of these cross sections is inherently difficult, as it requires isolating small quantities of pure LLFP material from spent nuclear fuel, which is both radiologically hazardous and logistically complex. Consequently, there is a scarcity of experimental data, particularly in the fast and epithermal neutron energy ranges. For example, the neutron capture cross section of Sn-126 has only two experimental data points, leaving the evaluated libraries (such as TENDL, JENDL, and ENDF) to rely almost entirely on theoretical model predictions \cite{ref:sn1} \cite{ref:sn2}. Some work exists on quantifying proton reaction uncertainties via cross-section perturbation, but only in the energy ranges of 8-30 MeV~\cite{ref:wickert}. This lack of experimental validation limits the accuracy of predicted transmutation rates and underscores the need for improved nuclear data to support future design and optimization of spallation-driven transmutation systems.

\subsection{FISPACT}
FISPACT is a computational tool and database system designed to simulate the behavior of materials undergoing nuclear depletion and transmutation reactions \cite{ref:fispact}. It is capable of both time-dependent and static analyses of irradiated isotopes, providing a detailed inventory of the resulting isotopes. The system is composed of three primary elements: the core software responsible for modeling depletion and transmutation, a set of nuclear data libraries, and rigorous validation and verification protocols. The nuclear data libraries are formatted in ENDF6, with users able to supplement these with additional custom libraries if needed. FISPACT processes inputs such as the particle flux spectrum, initial composition and count of target isotopes, irradiation duration, and particle type. Its outputs include data on isotopic activity, the composition of irradiated materials, radiotoxicity levels, and other essential parameters. For this work, results from PHITS simulations, including the flux-energy distribution and the quantity of irradiated atoms, are fed into FISPACT to perform a 0-dimensional transmutation rate analysis for LLFPs with the CCFE-709 group structure. FISPACT simulations used TENDL-2017 data for neutron reactions \cite{ref:tendl}, as well as HEIR-01 data for proton reactions \cite{ref:HEIR}.

\subsection{Target-Blanket Concept}
This work investigates optimal target–blanket material combinations to maximize transmutation efficiency. The general layout features a central spallation target surrounded by LLFP material and moderator material, along with an outer reflector surrounding the entire geometry.

The central target is 2 m in axial length, with a diameter of 10 cm chosen to ensure utilization the majority of incident protons while remaining thin enough to allow the secondary neutrons produced in spallation reactions to escape without significant absorption. A preliminary study on target width showed that an optimal target diameter was between 10 and 15 cm. The target also features a 1 cm in radius, 1 m long channel that the proton beam is incident upon. The point of the void region is for the beam to interact with the spallation target in the middle of the tank instead of on the outer surface of the spallation target. The beam travels through the void region, and interacts with the spallation target, leading to the previously mentioned desired neutron production. The material choice for the spallation target has a strong impact on neutron production and therefore transmutation rates, and multiple material choices were individually analyzed with the combination of PHITS and FISPACT. Further discussion of target material choice is continued in Section~\ref{sec-results-target}.

\begin{figure}[htbp]
    \centering
    \includegraphics[width=1.0\linewidth]{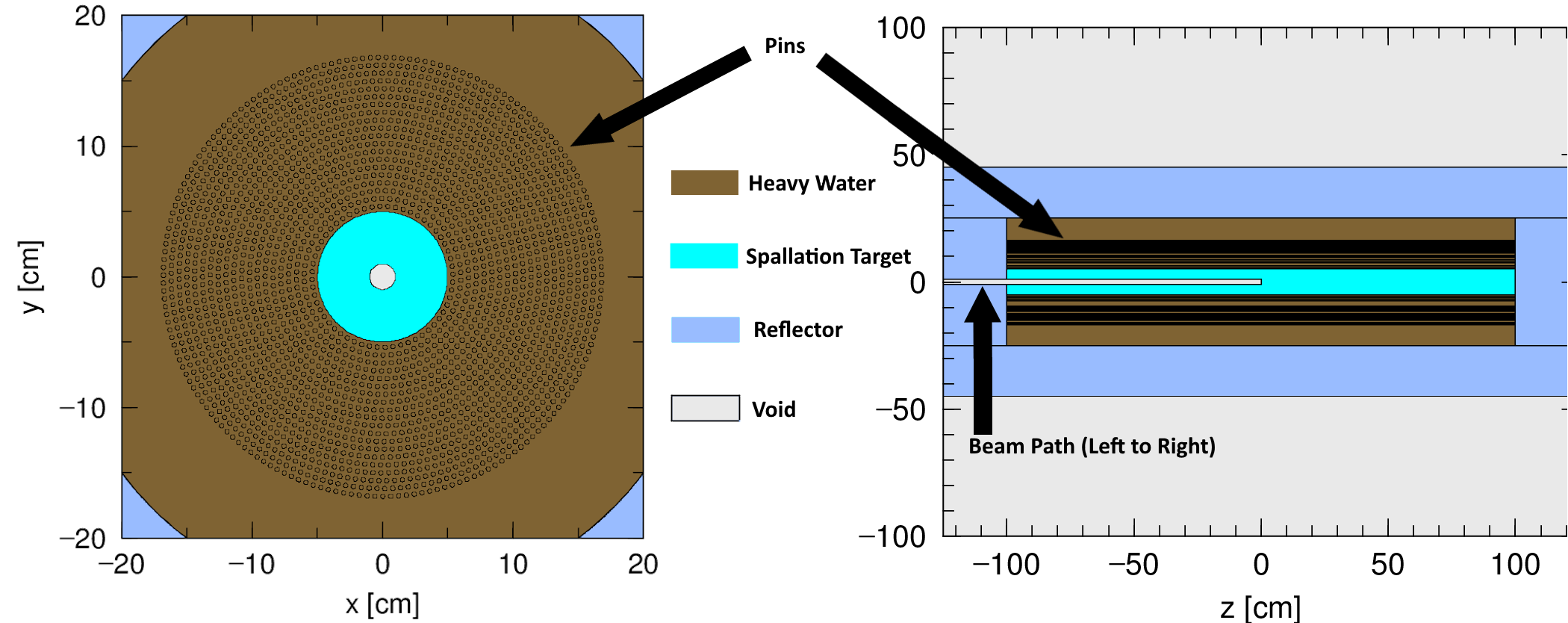}
    \caption{Geometry of tank layout for tallying neutron spectrum at different pin layers from a top down view (left) and a side view (right). Light blue represents the spallation target, white represents the beam, brown represents heavy water, and dark blue is the reflector. Note that the LLFP pins surround the spallation target and that the beam travels from left to right on the z-axis.}
    \label{fig:test}
\end{figure}

The LLFP material is arranged in cylindrical pins that surround the spallation target in a concentric pattern. These pins have a radius of 0.2 cm to avoid spatial self shielding. Each layer/pin edge is separated from each other with 0.2 cm of heavy water. The chosen moderator is heavy water (D$_2$O) to avoid neutron absorption while maintaining high thermalization rates. The target, surrounding LLFPs, and moderator are surrounded by a 25 cm thick beryllium reflector both axially and radially. Beryllium's high scattering cross-section as well as high (n,2n) cross-section makes the material a strong reflector for neutrons. The inner radius of this reflector is 1 m, and the total height of the geometry is 2.5 m. Figure \ref{fig:test} shows this test layout, where the pins represent where the LLFP pins surround the spallation target.

In this work, the LLFP mass in the system is limited to a value relative to the annual output of an operating commercial reactor, in order to identify an optimal system for transmuting a reasonable amount of each LLFP. The LLFP pin radii is set as 0.2 cm to take advantage of the higher efficiency of the initial mass percentage burned per year compared to the larger radii cases. In Table \ref{tab:Pinnum}, the annual mass output of a typical 3 GWth reactor is considered for each LLFP to determine how many layers and pins each LLFP requires \cite{ref:llfp}. The number of pins required are proportional to the relative abundance of the LLFP in spent fuel.

\begin{table}[ht]
    \centering
    \caption{Mass of LLFP produced per year in a 3 GWth reactor and the number of 2 m length, 0.2 cm radius rods required to represent these masses in their elemental composition. 
    The numbers of pins are slightly adjusted for each nuclide for rounding purposes and to accommodate the combined case described later.}
    \label{tab:Pinnum}
    \begin{tabular}{lccc}
        \toprule
        \textbf{Nuclide} & \textbf{Production Rate (kg/GW$_\mathrm{th}$/year) \cite{ref:llfp}} & \textbf{No. of Pins}\\
        \midrule
        Se-79  & 0.066  &  14 \\
        Zr-93  & 8.04  &  702\\
        Tc-99  & 8.54  &   82\\
        Sn-126 & 0.30  &  18\\
        I-129  & 1.96  &   66\\
        Cs-135 & 2.76  &  430\\
        \bottomrule
    \end{tabular}
\end{table}

Spallation neutrons are created in the very fast range (1 GeV is the upper range for neutrons simulated) and quickly moderate in the heavy water surrounding the target. This means the neutron flux spectrum near the target should be relatively fast, and that spectrum should soften as one moves outwards from the target. Taking this into account, it is preferable to put materials close to the target that feature higher neutron transmutation reaction rates in a fast spectrum.  Section~\ref{sec:res-sensitivity} shows the efforts to quantify each LLFP material's sensitivity to neutron spectrum thermalization.

\section{Results}

\subsection{Spallation Target}\label{sec-results-target}

The selection of the spallation target material is an important factor in determining the overall efficiency of the system, as the transmutation process depends on neutrons from the proton induced spallation reactions. Thus, it is advantageous to use neutron-rich, high-Z materials that yield higher neutron production rates. Lead and depleted uranium are therefore compared as candidate targets due to their high neutron yields. Tc-99 is chosen for these comparisons because it is the only LLFP present in spent fuel as a single isotope. Other LLFPs occur as mixtures of isotopes, which introduce additional reaction channels and reduce the overall efficiency of transmutation. The isotopic purity of Tc-99 makes it a clear benchmark for examining the effect of target material on neutron generation. Table \ref{tab:rate} showcases the neutrons produced per incoming proton for each of the three listed materials at various beam energies. The targets were chosen to be 10 cm diameter and 61 cm long cylinders to emulate experimental data and verify the results of the PHITS runs \cite{ref:carpenter}. 

\begin{table}[ht]
    \centering
    \caption{Spallation Rate (Neutrons created per incoming proton) comparison at differing energies of depleted uranium (U-238), lead, and Tc-99 10 cm diameter and 61 cm long cylinder targets. }
    \label{tab:rate}
    \begin{tabular}{ccccc}
        \toprule
        \textbf{Energy (MeV)}   & \textbf{Tc-99} & \textbf{Lead} & \textbf{U-238} \\
        \midrule
        500  & 5.166 $\pm$ 0.023 & 7.496 $\pm$ 0.034   & 14.363 $\pm$ 0.063\\
        750  & 9.237 $\pm$ 0.030 & 13.633 $\pm$ 0.042  & 26.308 $\pm$ 0.079\\
        1000  & 12.937 $\pm$ 0.034 & 19.171 $\pm$ 0.048  & 37.801 $\pm$ 0.087\\
        \bottomrule
    \end{tabular}
\end{table}

The results show a clear increase in neutron yield with both higher proton energy and heavier target material. Across all energies, U-238 exhibits the highest neutron production, followed by lead, with Tc-99 producing the lowest yields. Tc-99 being used as the spallation target comes with the advantage of additional transmutation of itself as the target. However, a preliminary test shows that a 10 cm diameter and 200 cm long Tc-99 cylinder (177.03kg) when exposed to a 1000 MeV proton beam will only result in 0.89 kg of additional transmutation in the target per year (around 0.5\% of the spallation target mass). In contrast, lead and depleted uranium produce substantially higher neutron yields from proton-induced spallation reactions, which would enhance transmutation in the surrounding LLFP regions at a more efficient rate than the incremental benefit gained from direct Tc-99 transmutation in the target. Proton-induced transmutation within the target itself is not the primary objective, as protons have shallow penetration depths and do not provide the same level of reaction multiplication as the neutrons generated in the spallation process. As can be seen in Table \ref{tab:rate}, the spallation rate for depleted uranium is higher than lead, implying that the depleted uranium is a better spallation target than lead. However, unlike lead, irradiating depleted uranium triggers fission reactions that generate more heat and new fission products. Thus, depleted uranium and lead are chosen as the spallation target candidates in this paper, and are compared against each other.

\subsection{Spectral Sensitivity Analysis}\label{sec:res-sensitivity}

Because the goal is to arrive at an optimal design where all the LLFP material is contained within a single tank for transmutation, there needs to be some thought put into how the materials are ordered. Neutrons leaving the spallation target will be predominantly fast, with energies upwards of 1 GeV, and will rapidly thermalize in the surrounding moderator. When ordering LLFP pins, the nuclide with the largest sensitivity to thermalization of neutrons should be placed the furthest away from the target, where neutron flux will be most thermal. The work of this subsection details quantifying each LLFP's spectral sensitivity in order to facilitate optimal ordering.

Preliminary runs without the LLFP pins and purely D$_2$O moderator show that there is no true fast region, as the neutrons are thermalized very quickly by the D$_2$O. The region closest to the target (lead in this case) had a neutron flux that was 60\% thermal (below 4 eV), and 10\% fast (greater than 1 MeV). At 11 cm from the target, the flux was 78\% thermal and only 3\% fast. A figure showing the neutron flux spectrum at different radial positions can be found in the supplementary information, Figure S1.

In order to sample the sensitivity of each LLFP to neutron flux spectrum, each LLFP's number of pins (specified in Table~\ref{tab:Pinnum} were placed surrounding the spallation target. These pins were then shifted radially outwards to 11 different layer positions, with 0.4 cm spacing between each layer. In cases where the number of LLFP pins exceeds the capacity of a single layer, the additional pins are placed in subsequent outer layers, meaning the case of "layer 0" means that the closest LLFP pins to the target were placed in layer 0.

\begin{figure}[!htbp]
    \centering
    \includegraphics[width=0.9\linewidth]{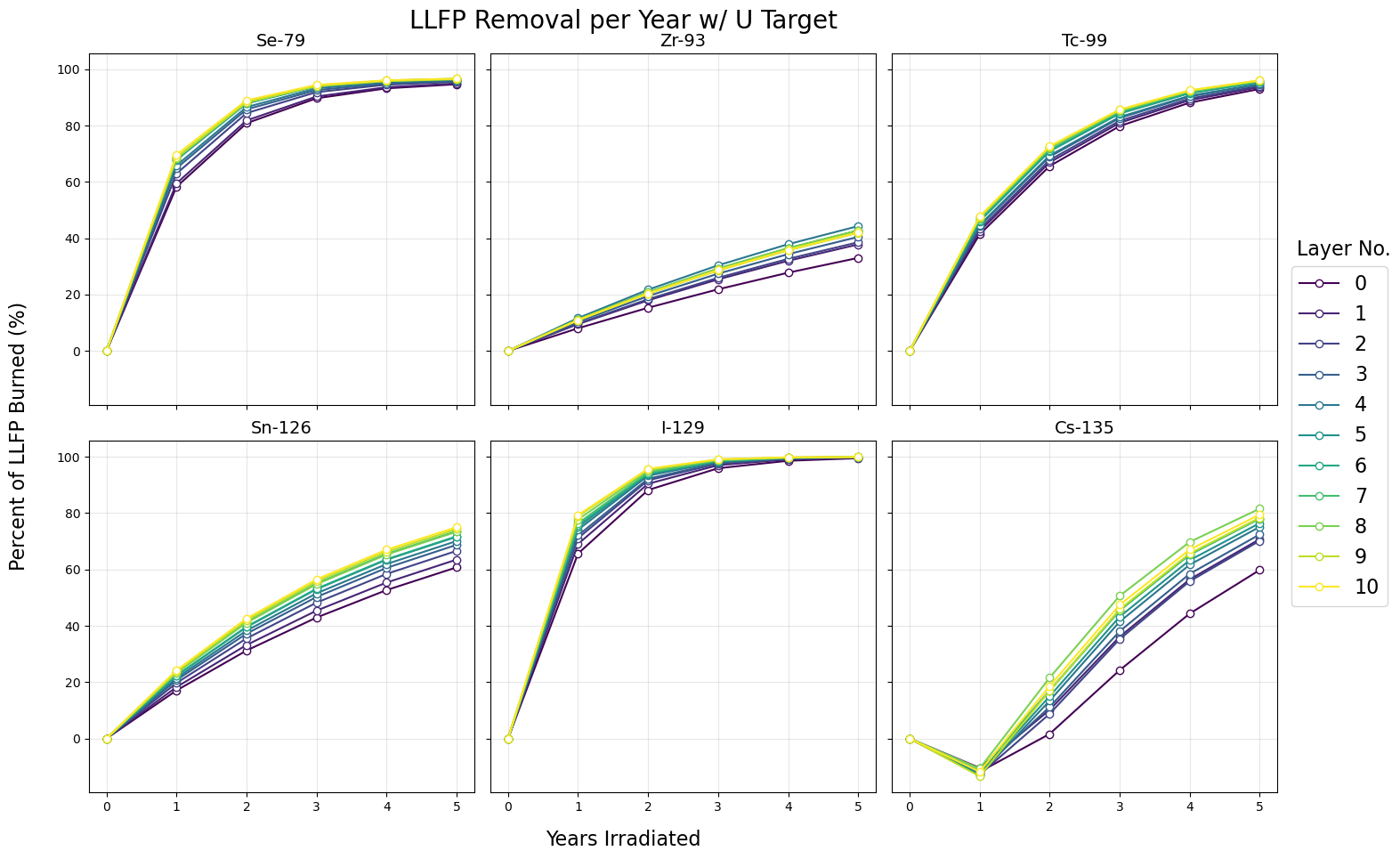}
    \caption{LLFP removal percent per year at different layer positions. Layer 0 is the layer closest to the spallation target (5.4 cm away from the center of the geometry. 0.4 cm away from the target edge), while layer 10 is the furthest (11.4 cm away from the center).} 
    \label{fig:spec-results}
\end{figure}

To calculate transmutation rates for each case, PHITS was used to calculate the average neutron flux in the LLFP pins created from the proton-based spallation, with FISPACT used to calculate transmutation rates from this flux. The results for percentage of LLFP removal per year for each material, specifically with a depleted uranium spallation target, can be seen in Figure \ref{fig:spec-results}.

For some of the nuclides, such as Se-79, Tc-99, and I-129, increased thermalization from being further from the source has a minor impact that mostly washes out by 5 years of irradiation. For others, such as Sn-126 and Cs-135, there is clear and significant increases in transmutation amounts due to a more thermal spectrum, in the range of 15-20\% increases in transmutation after 5 years. Zr-93 has more unclear response to the change in position, likely because the scattering cross-section dominates the total cross-section, along with some stochastic noise from the PHITS simulation.

Cs-135 is characterized by a net negative transmutation rate during the first 1-2 years of irradiation. This is due to Cs-133 and Cs-134 having larger capture cross-sections than Cs-135, particularly in the thermal range. This means that transmuting Cs-135, when contained with its sister isotopes, requires burning through the Cs-133 and Cs-134 first, which in this case took approximately 2 years of irradiation.

Comparisons of each LLFP's sensitivity to neutron spectrum can be estimated by looking at the relative difference in mass transmuted between layer 0 and layer 10, as seen in Table~\ref{tab:relative}. The LLFP with the strongest relative increase in mass transmuted can be inferred to be the most sensitive to a more thermal spectrum, and should be placed at the outside of the combined arrangement. As the cesium isotope is most sensitive of the LLFPs and zirconium has a low burn rate regardless of position, the optimal ordering from outwards-to-inwards was set to Cs$\rightarrow$Tc$\rightarrow$Se$\rightarrow$I$\rightarrow$Sn$\rightarrow$Zr. All elements besides cesium and zirconium have relatively few pins, and their specific ordering in the middle layer has little impact.

Note that the relative differences change depending on what target is used. This is because the depleted uranium target produces more neutrons per spallation (noted in Table~\ref{tab:rate}) as well as a significant amount of fission neutrons. These additional neutrons skew the spectrum in the moderator, making the region closest to the target faster, and making the difference in thermal proportion of flux larger as the layers move away from the target. Other impacts due to target choice are analyzed in Section~\ref{sec:target-impact}. Finally, note that because volume of the LLFP pins is constant in every layer case, the further radii cases have a lower total flux. This is due to the $1/r$ geometric attenuation in a cylindrical space and any leakage or absorption that occurs before neutrons reach the outer layer positions. For example, the layer 11 case with zirconium pins has 76.3\% of the total flux that the layer 0 case has. This means that while an estimation of true spectral sensitivity from these results would be slightly under-predictive, the sensitivity of one LLFP compared to another will remain consistent.

\begin{table}[ht]
\centering
\begin{tabular}{c c|c c c}
\hline
    \multirow{2}{*}{Target} & \multirow{2}{*}{Nuclide} & \multicolumn{2}{|c}{Mass Transmuted in} & \multirow{2}{*}{Relative Difference} \\
    & & Layer No. 0 & Layer No. 10 & \\
\hline
    \multirow{6}{*}{Pb} & Cs-135 & 4.196 & 4.801 & 7.3\% \\
    & Tc-99 & 21.33 & 22.51 & 4.6\% \\
    & Se-79 & 0.120 & 0.201 & 0.9\% \\
    & I-129 & 5.870 & 5.895 & 0.4\% \\
    & Sn-126 & 0.649 & 0.642 & 0.7\% \\ 
    & Zr-93 & 7.780 & 6.930 & -3.7\% \\
    \hline
    \multirow{6}{*}{U} & Cs-135 & 4.979 & 6.615 & 19.6\% \\
    & Sn-126 & 0.560 & 0.692 & 14.3\% \\
    & Zr-93 & 7.670 & 9.780 & 9.1\% \\
    & Tc-99 & 23.92 & 24.71 & 3.1\% \\
    & Se-79 & 0.198 & 0.203 & 2.0\% \\
    & I-129 & 5.881 & 5.908 & 0.5\% \\
    \hline
\end{tabular}
\caption{Relative difference in LLFP mass transmuted due to changing layer position. Masses are in kg, and are the results from 5 years of irradiation. Nuclides are ordered from highest to lowest relative difference, for each respective target material.}
\label{tab:relative}
\end{table}

\subsection{Combined Geometry}

\begin{figure}[!htbp]
    \centering
    \includegraphics[width=1.0\linewidth]{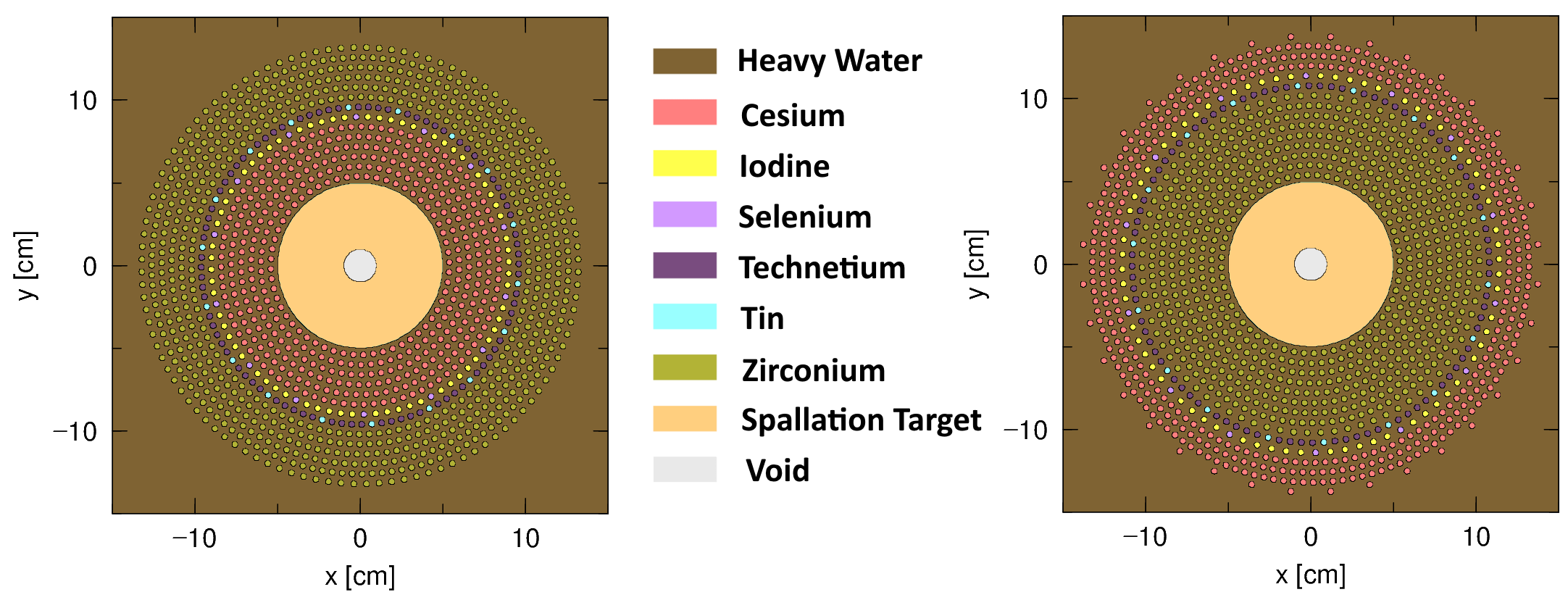}
    \caption{Combined test layouts for both pin orderings. On the left is cesium first while on the right is zirconium first.}
    \label{fig:combined}
\end{figure}

With element ordering determined by the estimation of spectral sensitivity, the PHITS/FISPACT analysis can be extended to the geometry that incorporates all LLFPs generated from one year of PWR operation within the transmutation tank. As this is an optimization study, all the LLFPs are combined into one tank to increase the transmutation rate as much as possible. The cases were run with both the order as described at the end of Section~\ref{sec:res-sensitivity}, as well as the reverse order for illustration purposes. These combinations can be seen in Figure~\ref{fig:combined}.  

\begin{figure}[!htbp]
    \centering
    \includegraphics[width=0.9\linewidth]{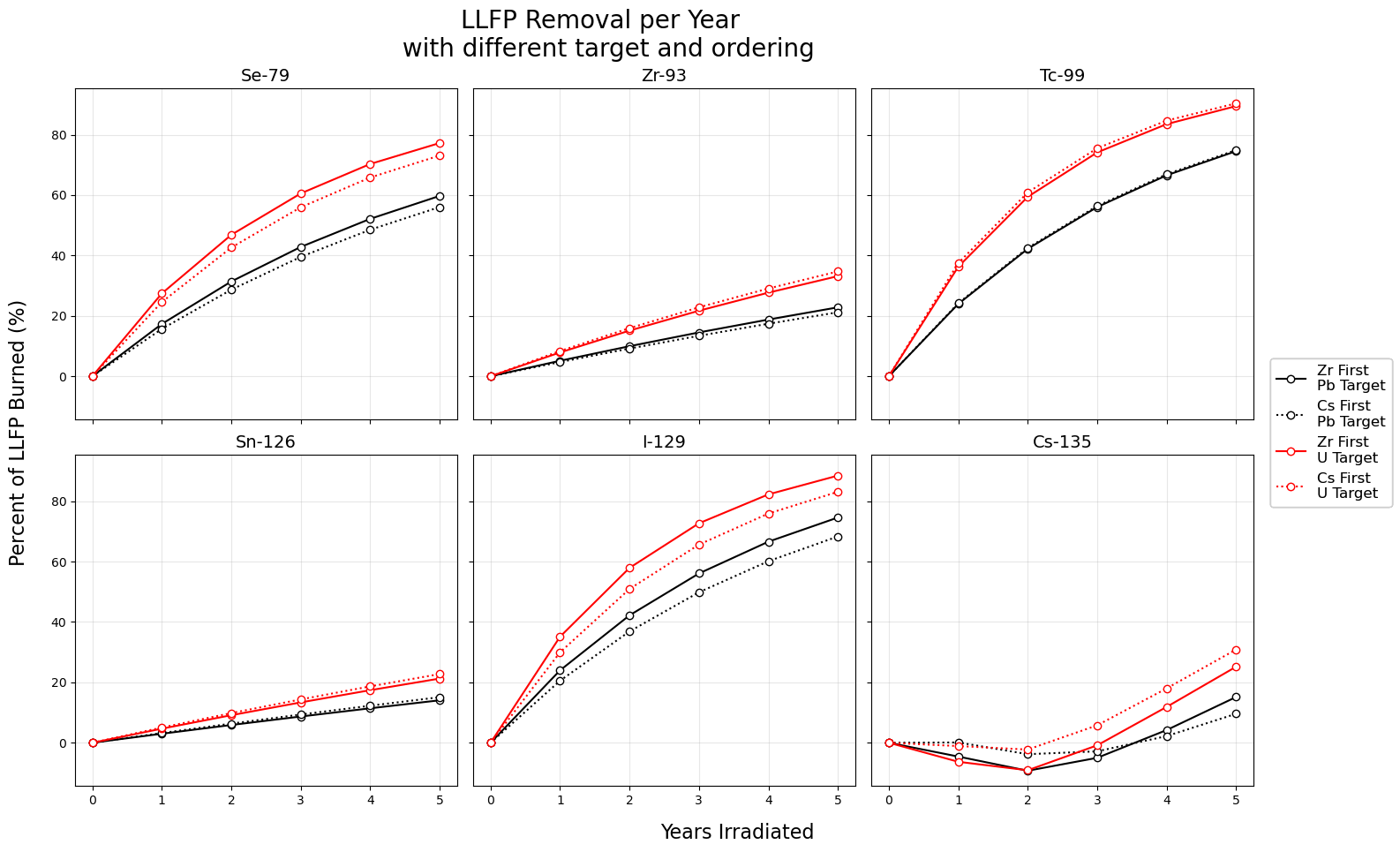}
    \caption{Combined results for both ordering and target choice for all LLFPs. Results are in percentage of LLFP transmuted versus year.}
    \label{fig:combined_res}
\end{figure}

After performing the PHITS and FISPACT simulation, some trends can be highlighted from the transmutation rates of each permutation of pin ordering and target choice. These rates of each permutation of target material and ordering can be seen in Figure~\ref{fig:combined_res}. 

The effect of elemental ordering is varied across the different LLFPs. For the low-pin-number nuclides (Se-79, Tc-99, Sn-126, I-129), the ordering either doesn't have a discernible impact, or there is a minor preference for the combination with Zr closest to the target. This is likely due to the aforementioned transparency of zirconium, allowing a higher magnitude of flux to reach the inner LLFPs. This transparency also results in Zr-93 having relatively little sensitivity to ordering. 

Cs-135, on the other hand, has an interesting behavior of having a higher rate close to the target when using uranium, yet a higher rate when away from the target when using lead. This is due to the competing balance of flux magnitude versus flux spectrum. For the lead target, the increased capture from neutron thermalization outweighs the loss of total flux due to attenuation and capture in other LLFPs. For the uranium target that has a higher neutron-per-proton rate, the increased flux near the target outweighs the gains from thermalization at the outer radius. Additionally, the depleted uranium target's fission spectrum neutrons have less moderation necessary to thermalize compared to solely spallation neutrons. Also note how the rate for all combinations is now lowered to such a level that it takes between 3 to 4 years to reach positive net transmutation of Cs-135. This is an effect purely of flux magnitude being lower due to the other elements being present in the geometry. This result is enough to suggest that cesium should likely be transmuted on its own, because paying for the accelerated proton beam to be active for multiple years before any positive reduction in Cs-135 levels is likely untenable for a prospective user. 

Across all nuclides, the depleted uranium target results in a 10-25\% increase in burn rate, which can be predominantly attributed to a higher magnitude of neutron flux. There are other considerations to take into account when using uranium as a target, however, which will be discussed in the following section.

Analysis of the impact of the irradiation with different target materials and ordering on radiotoxicity of the LLFP materials can be found in the Supplementary materials, specifically Figures S2, S3, and Table S6. Radiotoxicity here is defined as the ingestion dose  retrieved from the FISPACT output, which corresponds to ICRP dose coefficients~\cite{ref:icrp}.

\subsection*{Impact of Target Material}\label{sec:target-impact}

 Depleted uranium has been a studied spallation target given its high neutron density and therefore higher number of neutrons produced per proton-based spallation \cite{ref:carpenter}.  More neutrons will lead to more LLFP removal in the set up, minus some spectral difference in neutrons produced from the target. 
 
 To generalize the results in Figure \ref{fig:spec-results} and Table~\ref{tab:relative}, the use of the depleted uranium target causes an increase in transmutation for every studied nuclide when measured at the outer layer position. After 5 years of simulated transmutation, these increases range from 40\% for Cs-135 to ~0\% for Se-79 and I-129. This difference is at its greatest in the first year and diminishes over time. This trend continues for the combined geometry results in Figure~\ref{fig:combined_res}. The increased neutrons resulting from the uranium target drive increases in transmutation rate ranging from 5\% (Sn-126) to 20\% (Se-79).

 To this point, this analysis is missing a large component; the uranium target has a non-negligible fission rate that both produces neutrons \textit{and} LLFP nuclides. If the uranium target’s production of LLFPs outweighs the increased transmutation from more spallation neutrons, then changing targets would be inadvisable. To estimate the LLFP generation in the depleted uranium target, PHITS-generated proton and neutron flux in the target was used as an input for two FISPACT runs. The neutron run considered fission due to neutrons in the target, and the proton run used TENDL incident proton data for fissions, as well as HEIR high-energy cascade data for fission and evaporation modeling. 
 
 The results show a combined-particle fission rate of 6.87e+17 (1/sec), or 3.67 fissions per source proton, and 0.8\% of these fissions are from the source protons. Protons account for roughly 4.5\% of all LLFP generation in the uranium target. This results in LLFP generation rates ranging between 0.33 kg per year of Cs-135, and 2.2 g per year of Se-79. Comparing these rates to the PWR generation rates in Table~\ref{tab:Pinnum}, Cs-135 is the highest LLFP generated, with the target producing roughly 4\% of what a 3 GW$_\text{th}$ PWR produces in a year, assuming the target is irradiated for a year as well. These uranium target generation rates are included with the transmutation rates, which can be seen in Table~\ref{tab:ordering2}. Note how the production of Cs-135 in the uranium target results in lead being the optimal target for Cs-135 transmutation.

\begin{table}[ht]
\centering
\renewcommand{\arraystretch}{1.2}
\begin{tabular}{c | c | c c c}
\hline
Target & Ordering & Nuclide & Average Removal Rate (kg/5yr) & Rel. Mass Removal Rate (\%/5yr) \\
\hline

\multirow{12}{*}{Pb} 
  & \multirow{6}{*}{Zr 1st} & Se-79  & 0.13  & 59.72 \\
  &                         & Zr-93  & 5.30   & 22.82 \\
  &                         & Tc-99  & 19.17 & 74.55 \\
  &                         & Sn-126 & 0.13  & 14.06 \\
  &                         & I-129  & 4.41    & 74.59 \\
  &                         & Cs-135 & 1.26  & 15.15 \\ \cline{2-5}

  & \multirow{6}{*}{Cs 1st} & Se-79  & 0.12  & 56.03 \\
  &                         & Zr-93  & 4.92    & 21.18 \\
  &                         & Tc-99  & 17.75   & 74.89 \\
  &                         & Sn-126 & 0.14  & 15.09 \\
  &                         & I-129  & 4.04  & 68.32 \\
  &                         & Cs-135 & 0.76  & 9.59 \\

\hline
\multirow{12}{*}{U} 
  & \multirow{6}{*}{Zr 1st} & Se-79  & 0.15  & 72.05 \\
  &                         & Zr-93  & 6.89 & 29.64 \\
  &                         & Tc-99  & 21.92 & 85.23 \\
  &                         & Sn-126 & 0.09  & 9.89 \\
  &                         & I-129  & 4.88  & 82.56 \\
  &                         & Cs-135 & 0.42   & 5.07 \\ \cline{2-5}

  & \multirow{6}{*}{Cs 1st} & Se-79  & 0.14  & 67.92 \\
  &                         & Zr-93  & 7.24 & 31.19 \\
  &                         & Tc-99  & 20.33   & 85.78 \\
  &                         & Sn-126 & 0.11  & 11.39 \\
  &                         & I-129  & 4.56  & 77.19 \\
  &                         & Cs-135 & 0.78 & 9.79  \\

\hline
\end{tabular}
\caption{Average and relative mass removal rates for nuclides under Pb and U targets. Note that the average removal rate takes into account the LLFP generation inside the depleted uranium target.}
\label{tab:ordering2}
\end{table}

One additional note is that 6.87e+17 fissions per second, and assuming 200 MeV per fission that deposits locally, this would translate to 22 MW of thermal power being deposited in the target. Assuming all protons have their kinetic energy locally, there would be 30 MW of thermal power deposited by the beam alone. Fission is a significant heat source that would require additional appropriate cooling of the target, or at the very least adequate thermal modeling of the target.

\subsection{Economic Analysis}

Because of the decision to limit the mass of each material based on a single PWR’s yearly generation, it makes sense to focus economic analysis of this facility around a single PWR-coupled process. For instance, a case where transmutation occurs on site and the accelerator is powered by electricity also generated on site. This means the cost of transmutation is based on the lost revenue from the electricity diverted for the accelerator rather than sold to the market (O\&M costs, capital costs, and reprocessing costs are ignored, although they would likely be substantial).  

Given the inputs of the beam having 30 mA current, 1 GeV protons, and an assumed 30\% accelerator power to grid efficiency \cite{ref:kovach}, operating the proton accelerator will require roughly 100 MWe, which aligns at around 10\% of the 3 GWth PWR’s generation capability. One way to estimate this economic impact is that 10\% of every watt-hour of energy can be considered “already spoken for”, so measures of cost based on electricity production will increase by a reciprocal 11.1\%. For example, an Nth-of-a-kind AP1000 would have its unsubsidized lifetime levelized cost of electricity (LCOE) increase from \$66/MWhre to \$81.5/MWhre, per data from Shirvan 2024 \cite{ref:Shirvan}.  

For quantifying cost of the transmutation process using lost revenue that would have come from selling on the wholesale energy market, a sale price of \$39.5/MWhre is assumed, based on the 2024 average real-time price for wholesale power in New England \cite{isone}. Based on 100 MWe required for the accelerator and an assumed 100\% capacity factor (to align with transmutation being continuous in the simulations), the lost revenue would be roughly \$34.6 million per year. This value can be combined with the 5-year averaged transmutation rates to estimate costs per LLFP, as seen in Table \ref{tab:economics}.

Also included in Table \ref{tab:economics} is a simple projection using the 5-year averaged data and assuming linear relationships between accelerator power and transmutation rate. This projection is the power required to burn 10\% of the LLFP per year (and its associated cost to PWR revenue). The assumptions for these projections make the values closer to estimations of order of magnitude; an accurate prediction of required power and cost would require more detailed simulation. Furthermore, Cs-135 has an initial period of production instead of generation, so using a linear assumption based on 5-year averaged data will cause over-prediction when estimating the power required to reach higher percentage burn-ups. 

The assumption on LLFP mass here is based on a single 3 GWth PWR’s yearly production. Better efficiencies and therefore economic projections are likely possible by combining 2-4 reactors’ LLFP production and only transmuting Cs-135, Zr-93, or a mix of the other four LLFPs (given their low pin counts). Most nuclear plants have multiple reactors on site, so the cost-benefit of using generated electricity instead of purchasing electricity would remain. 

\begin{table}[!ht]
\centering
\begin{tabular}{c |c | c c c c c}
\hline
    \multirow{2}{*}{Target} & \multirow{2}{*}{Ordering} & \multirow{2}{*}{Nuclide} &  Cost per kg & Cost per pct. & Power req. & ...Associated \\
   & & & (\$M/kg) & (\$M/\%) & for 10\%/yr (MW)& lost revenue (\$M) \\ \hline
    \multirow{12}{*}{Pb} & \multirow{6}{*}{Zr 1st} & Se-79 & 1381.4 & 2.9 & 83.7 & 29.0\\
    & & Zr-93 & 32.6 & 7.6 & 219.2 & 75.8\\
    & & Tc-99 & 9.0 & 2.3 &67.1 &23.2 \\
    & & Sn-126 & 1335.0&12.3 &355.7 &123.1 \\
    & & I-129 & 39.2&2.3 & 67.0 &23.2 \\
    & & Cs-135 & 137.2&11.4 &330.1 &114.2 \\ \cline{2-7}
    & \multirow{6}{*}{Cs 1st} & Se-79 & 1470.4&3.1 &89.2 &30.9 \\
    & & Zr-93 & 35.2&8.2 &236.1 &81.7 \\
    & & Tc-99 & 9.7&2.3 &66.8 &23.1 \\
    & & Sn-126 & 1242.0&11.5 &331.3 &114.6 \\
    & & I-129 & 42.8 &2.5 &73.2 &25.3 \\
    & & Cs-135 & 227.3& 18.0& 521.6&180.5 \\ \hline
    \multirow{12}{*}{U} & \multirow{6}{*}{Zr 1st} & Se-79 & 1145.1&2.4 &69.4 &24.0 \\
    & & Zr-93 & 25.1&5.8 &168.7 &58.4 \\
    & & Tc-99 & 7.9&2.0 &58.7 &20.3 \\
    & & Sn-126 & 1898.3&17.5 &505.8 &175.0 \\
    & & I-129 & 35.4&2.1 &60.6 &21.0 \\
    & & Cs-135 & 409.6&34.1 &985.6 &341.0 \\ \cline{2-7}
    & \multirow{6}{*}{Cs 1st} & Se-79 & 1212.9& 2.5&73.6 &25.4 \\
    & & Zr-93 & 23.9&5.5 &160.3 &55.5 \\
    & & Tc-99 & 8.5&2.0 &58.3 &20.2 \\
    & & Sn-126 & 1645.5&15.2 &438.9 &151.9 \\
    & & I-129 &37.9 &2.2 &64.8 &22.4 \\
    & & Cs-135 & 222.5& 17.7&510.5 &176.7 \\
    \hline
\end{tabular}
\caption{Relative difference in LLFP mass transmuted due to changing layer position. Masses are in kg. Nuclides are ordered from highest to lowest relative difference, for each respective target material.}
\label{tab:economics}
\end{table}

\section{Discussion}

This work has investigated the feasibility of using proton-induced spallation as a driver for the transmutation of long-lived fission products (LLFPs), focusing on the interplay between spallation target material, spectral sensitivity of individual nuclides, and the economics of accelerator operation. Through PHITS and FISPACT simulations, neutron yield characteristics were quantified for candidate spallation targets, showing that depleted uranium produces up to ~2× higher neutron yields than lead across the studied energies. This neutron gain translated to a 10–25\% increase in transmutation rates across most LLFPs, although it also introduced secondary LLFP generation and substantial thermal power deposition (~22 MW) in the target, highlighting important engineering trade-offs.

Nuclide specific behavior was found to strongly shape system efficiency. Zr-93 is costly to transmute due to its neutron transparency, with little improvement from spectral tailoring, while Cs-135 requires maximum neutron availability since Cs-133/134 must first be consumed before net reduction occurs. By contrast, Se-79, Tc-99, and I-129 transmute relatively easily, and Sn-126, though more resistant, still benefits from thermal flux. Most of the nuclides perform better under thermalized spectra, with Cs-135 and Sn-126 showing the greatest gains. When combined in a single geometry, the lower-mass nuclides can be efficiently co-transmuted, but including Cs-135 and Zr-93 reduces system effectiveness, suggesting they should be treated separately. From an economic perspective, coupling a 1 GeV, 30 mA accelerator to a 3 GWth PWR requires ~100 MWe, raising LCOE by ~11\% and costing ~\$34.6M/year in lost electricity sales. Costs per kilogram vary widely, with Tc-99 the most favorable (~\$9M/kg) and Cs-135 and Zr-93 the least due to their neutron demands.


Some limitations of this work are as follows: the spallation sources is assumed constant for 5 years for all of the FISPACT runs, and cross-section data does not have uncertainty included in the results, something that should be clarified in future work. There is also a lack of experimental neutron cross section data for the LLFPs. This work provides a systematic assessment that couples spallation target comparisons with spectral sensitivity based LLFP ordering within a mass constrained geometry representative of reactor scale waste inventories. The results highlight clear trade-offs between neutron yield, target material side-effects, isotope placement, and achievable burn rates. This framework provides a more realistic basis for assessing transmutation strategies and can guide future research in three directions: (i) experimental validation of optimized target–blanket designs, (ii) refinement of economic models including reprocessing and capital costs, and (iii) exploration of hybrid systems where nuclides such as Cs-135 are treated separately from the combined geometry. By integrating these aspects, accelerator-driven transmutation could evolve into a technically and economically credible approach for reducing the long-term radiotoxic burden of nuclear waste.
 \newpage

\bibliography{main}

\begin{thebibliography}{10}
\urlstyle{rm}
\expandafter\ifx\csname url\endcsname\relax
  \def\url#1{\texttt{#1}}\fi
\expandafter\ifx\csname urlprefix\endcsname\relax\def\urlprefix{URL }\fi
\expandafter\ifx\csname doiprefix\endcsname\relax\def\doiprefix{DOI: }\fi
\providecommand{\bibinfo}[2]{#2}
\providecommand{\eprint}[2][]{\url{#2}}

\bibitem{ref:alvarez}
\bibinfo{author}{Alvarez, R.}
\newblock \bibinfo{journal}{\bibinfo{title}{Spent nuclear fuel pools in the us}}.
\newblock {\emph{\JournalTitle{Institute for Policy Studies}}}  (\bibinfo{year}{2011}).

\bibitem{ref:llfp}
\bibinfo{author}{Yang, W.~S.}, \bibinfo{author}{Kim, Y.}, \bibinfo{author}{Hill, R.~N.}, \bibinfo{author}{Taiwo, T.~A.} \& \bibinfo{author}{Khalil, H.~S.}
\newblock \bibinfo{journal}{\bibinfo{title}{Long-lived fission product transmutation studies}}.
\newblock {\emph{\JournalTitle{Nuclear Science and Engineering}}} \textbf{\bibinfo{volume}{146}}, \bibinfo{pages}{291--318}, \doiprefix\url{10.13182/NSE04-A2411} (\bibinfo{year}{2004}).
\newblock \eprint{https://doi.org/10.13182/NSE04-A2411}.

\bibitem{ref:wigeland}
\bibinfo{author}{Wigeland, R.}
\newblock \bibinfo{journal}{\bibinfo{title}{Nuclear fuel cycle evaluation and screening – final report}}.
\newblock {\emph{\JournalTitle{Idaho National Laboratory Report}}} \doiprefix\url{No. INL/EXT-14-31465} (\bibinfo{year}{2014}).

\bibitem{ref:spal2}
\bibinfo{author}{Anderson, I.} \emph{et~al.}
\newblock \bibinfo{journal}{\bibinfo{title}{Research opportunities with compact accelerator-driven neutron sources}}.
\newblock {\emph{\JournalTitle{Physics Reports}}} \textbf{\bibinfo{volume}{654}}, \bibinfo{pages}{1--58}, \doiprefix\url{https://doi.org/10.1016/j.physrep.2016.07.007} (\bibinfo{year}{2016}).
\newblock \bibinfo{note}{Research opportunities with compact accelerator-driven neutron sources}.

\bibitem{ref:gary}
\bibinfo{author}{Russell, G.}
\newblock \bibinfo{title}{{Spallation Physics - An Overview}}.
\newblock In \emph{\bibinfo{booktitle}{ICANS-XI International Collaboration on Advanced Neutron Sources}} (\bibinfo{year}{1990}).

\bibitem{ref:transmut}
 \emph{\bibinfo{title}{Implications of Partitioning and Transmutation in Radioactive Waste Management}}.
\newblock No. \bibinfo{number}{435} in \bibinfo{series}{Technical Reports Series} (\bibinfo{publisher}{INTERNATIONAL ATOMIC ENERGY AGENCY}, \bibinfo{address}{Vienna}, \bibinfo{year}{2004}).

\bibitem{ref:physor}
\bibinfo{author}{Wickert, C.} \emph{et~al.}
\newblock \bibinfo{title}{Progress on transport and transmutation analysis of proton beam irradiating long-lived spent nuclear waste}.
\newblock In \emph{\bibinfo{booktitle}{International Conference on Physics of Reactors (PHYSOR 2024)}}, \bibinfo{pages}{2330--2337} (\bibinfo{organization}{ANS}, \bibinfo{address}{San Francisco, CA}, \bibinfo{year}{2024}).

\bibitem{ref:Kang}
\bibinfo{author}{Kang, C.~M.}, \bibinfo{author}{Kim, J.-K.} \& \bibinfo{author}{Kang, W.-G.}
\newblock \bibinfo{journal}{\bibinfo{title}{Nuclear transmutation of 99tc utilizing proton spallation and compact subcritical assembly}}.
\newblock {\emph{\JournalTitle{Nuclear Technology}}} \textbf{\bibinfo{volume}{211}}, \bibinfo{pages}{1337--1346}, \doiprefix\url{10.1080/00295450.2024.2387409} (\bibinfo{year}{2025}).
\newblock \eprint{https://doi.org/10.1080/00295450.2024.2387409}.

\bibitem{ref:Chiba}
\bibinfo{author}{Chiba, S.} \emph{et~al.}
\newblock \bibinfo{journal}{\bibinfo{title}{Method to reduce long-lived fission products by nuclear transmutations with fast spectrum reactors}}.
\newblock {\emph{\JournalTitle{Scientific Reports}}} \textbf{\bibinfo{volume}{7}}, \doiprefix\url{10.1038/s41598-017-14319-7} (\bibinfo{year}{2017}).

\bibitem{ref:anes}
\bibinfo{author}{Sun, X.~Y.} \emph{et~al.}
\newblock \bibinfo{journal}{\bibinfo{title}{Transmutation of long-lived fission products in an advanced nuclear energy system}}.
\newblock {\emph{\JournalTitle{Scientific Reports}}} \textbf{\bibinfo{volume}{12}}, \doiprefix\url{10.1038/s41598-022-06344-y} (\bibinfo{year}{2022}).

\bibitem{ref:Jin}
\bibinfo{author}{Jin, M.-T.}, \bibinfo{author}{Xu, S.-Y.}, \bibinfo{author}{Yang, G.-M.} \& \bibinfo{author}{Su, J.}
\newblock \bibinfo{journal}{\bibinfo{title}{Yield of long-lived fission product transmutation using proton-, deuteron-, and alpha particle-induced spallation}}.
\newblock {\emph{\JournalTitle{Nuclear Science and Techniques}}} \textbf{\bibinfo{volume}{32}}, \doiprefix\url{10.1007/s41365-021-00933-8} (\bibinfo{year}{2021}).

\bibitem{ref:phits}
\bibinfo{author}{Sato, T.} \emph{et~al.}
\newblock \bibinfo{journal}{\bibinfo{title}{Features of particle and heavy ion transport code system (phits) version 3.02}}.
\newblock {\emph{\JournalTitle{Journal of Nuclear Science and Technology}}} \textbf{\bibinfo{volume}{55}}, \doiprefix\url{10.1080/00223131.2017.1419890} (\bibinfo{year}{2018}).

\bibitem{ref:INCL}
\bibinfo{author}{Boudard, A.}, \bibinfo{author}{Cugnon, J.}, \bibinfo{author}{Leray, S.} \& \bibinfo{author}{Volant, C.}
\newblock \bibinfo{journal}{\bibinfo{title}{Intranuclear cascade model for a comprehensive description of spallation reaction data}}.
\newblock {\emph{\JournalTitle{Physical Review C}}} \textbf{\bibinfo{volume}{66}}, \bibinfo{pages}{044615} (\bibinfo{year}{2002}).

\bibitem{ref:GEM}
\bibinfo{author}{Furihata, S.}
\newblock \bibinfo{journal}{\bibinfo{title}{Statistical analysis of light fragment production from medium energy proton-induced reactions}}.
\newblock {\emph{\JournalTitle{Nuclear Instruments and Methods in Physics Research Section B: Beam Interactions with Materials and Atoms}}} \textbf{\bibinfo{volume}{171}}, \bibinfo{pages}{251--258} (\bibinfo{year}{2000}).

\bibitem{ref:JENDL}
\bibinfo{author}{Shibata, K.} \emph{et~al.}
\newblock \bibinfo{journal}{\bibinfo{title}{Jendl-4.0: a new library for nuclear science and engineering}}.
\newblock {\emph{\JournalTitle{Journal of Nuclear Science and Technology}}} \textbf{\bibinfo{volume}{48}}, \bibinfo{pages}{1--30} (\bibinfo{year}{2011}).

\bibitem{ref:sn1}
\bibinfo{author}{Manning, B.} \emph{et~al.}
\newblock \bibinfo{journal}{\bibinfo{title}{Informing direct neutron capture on tin isotopes near the $n=82$ shell closure}}.
\newblock {\emph{\JournalTitle{Phys. Rev. C}}} \textbf{\bibinfo{volume}{99}}, \bibinfo{pages}{041302}, \doiprefix\url{10.1103/PhysRevC.99.041302} (\bibinfo{year}{2019}).

\bibitem{ref:sn2}
\bibinfo{author}{Sheng-dong, Z.}
\newblock \bibinfo{journal}{\bibinfo{title}{Progress in nuclear data measurement of long-lived fission products}}.
\newblock {\emph{\JournalTitle{Atomic Energy Science and Technology}}} \textbf{\bibinfo{volume}{40}}, \bibinfo{pages}{199--205}, \doiprefix\url{10.7538/yzk.2006.40.02.0199} (\bibinfo{year}{2006}).

\bibitem{ref:wickert}
\bibinfo{author}{Wickert, C.~I.}
\newblock \emph{\bibinfo{title}{Non-Neutron Transmutation of Spent Fuel}}.
\newblock Master's thesis, \bibinfo{school}{Massachusetts Institute of Technology} (\bibinfo{year}{2024}).

\bibitem{ref:fispact}
\bibinfo{author}{Sublet, J.-C.} \emph{et~al.}
\newblock \bibinfo{journal}{\bibinfo{title}{{FISPACT-II}: An advanced simulation and inventory code for nuclear research and waste management}}.
\newblock {\emph{\JournalTitle{Nuclear Data Sheets}}} \textbf{\bibinfo{volume}{148}}, \bibinfo{pages}{305--310}, \doiprefix\url{10.1016/j.nds.2018.02.022} (\bibinfo{year}{2018}).

\bibitem{ref:tendl}
\bibinfo{author}{Koning, A.} \emph{et~al.}
\newblock \bibinfo{journal}{\bibinfo{title}{Tendl: Complete nuclear data library for innovative nuclear science and technology}}.
\newblock {\emph{\JournalTitle{Nuclear Data Sheets}}} \textbf{\bibinfo{volume}{155}}, \bibinfo{pages}{1--55}, \doiprefix\url{https://doi.org/10.1016/j.nds.2019.01.002} (\bibinfo{year}{2019}).
\newblock \bibinfo{note}{Special Issue on Nuclear Reaction Data}.

\bibitem{ref:HEIR}
\bibinfo{author}{Fleming, M.}, \bibinfo{author}{Eastwood, J.}, \bibinfo{author}{Stainer, T.}, \bibinfo{author}{David, J.-C.} \& \bibinfo{author}{Mancusi, D.}
\newblock \bibinfo{journal}{\bibinfo{title}{Heir: A high-energy intra-nuclear cascade liège-based residual nuclear data library for simulation with fispact-ii}}.
\newblock {\emph{\JournalTitle{Nuclear Instruments and Methods in Physics Research Section A: Accelerators, Spectrometers, Detectors and Associated Equipment}}} \textbf{\bibinfo{volume}{908}}, \bibinfo{pages}{291--297}, \doiprefix\url{https://doi.org/10.1016/j.nima.2018.06.065} (\bibinfo{year}{2018}).

\bibitem{ref:carpenter}
\bibinfo{author}{Carpenter, J.~M.} \& \bibinfo{author}{Yelon, W.~B.}
\newblock \bibinfo{title}{2. neutron sources}.
\newblock In \bibinfo{editor}{Sköld, K.} \& \bibinfo{editor}{Price, D.~L.} (eds.) \emph{\bibinfo{booktitle}{Neutron Scattering}}, vol.~\bibinfo{volume}{23} of \emph{\bibinfo{series}{Methods in Experimental Physics}}, \bibinfo{pages}{99--196}, \doiprefix\url{https://doi.org/10.1016/S0076-695X(08)60555-4} (\bibinfo{publisher}{Academic Press}, \bibinfo{year}{1986}).

\bibitem{ref:icrp}
\bibinfo{author}{Eckerman, K.}, \bibinfo{author}{Harrison, J.}, \bibinfo{author}{Menzel, H.}, \bibinfo{author}{Clement, C.} \emph{et~al.}
\newblock \bibinfo{journal}{\bibinfo{title}{Icrp publication 119: compendium of dose coefficients based on icrp publication 60}}.
\newblock {\emph{\JournalTitle{Annals of the ICRP}}} \textbf{\bibinfo{volume}{41}}, \bibinfo{pages}{1--130} (\bibinfo{year}{2012}).

\bibitem{ref:kovach}
\bibinfo{author}{Kovach, A.}, \bibinfo{author}{Parfenova, A.}, \bibinfo{author}{Grillenberger, J.} \& \bibinfo{author}{Seidel, M.}
\newblock \bibinfo{journal}{\bibinfo{title}{Energy efficiency and saving potential analysis of the high intensity proton accelerator hipa at psi}}.
\newblock {\emph{\JournalTitle{Journal of Physics: Conference Series}}} \textbf{\bibinfo{volume}{874}}, \bibinfo{pages}{012058}, \doiprefix\url{10.1088/1742-6596/874/1/012058} (\bibinfo{year}{2017}).

\bibitem{ref:Shirvan}
\bibinfo{author}{Shirvan, K.}
\newblock \bibinfo{journal}{\bibinfo{title}{{2024 total cost projection of Next AP1000}}}.
\newblock {\emph{\JournalTitle{MIT CANES}}}  (\bibinfo{year}{2024}).

\bibitem{isone}
\bibinfo{author}{{ISO New England}}.
\newblock \bibinfo{title}{{Key Grid and Market Stats}}.
\newblock \bibinfo{howpublished}{\url{https://www.iso-ne.com/about/key-stats/markets}}.
\newblock \bibinfo{note}{Accessed: 2025-09-22}.

\end{thebibliography}

\section*{Author contributions statement}

The idea behind the work was conceived by A.D. and B.F.. G.T. implemented the PHITS analysis and W.R.K. implemented the FISPACT analysis. J.Y. helped prepare scripts for PHITS geometries. A.D. and B.F. advised G.T. and W.R.K. throughout the work. G.T. and W.R.K. analyzed the results. G.T. and W.R.K. wrote the manuscript text, A.D. and B.F. revised the manuscript text. 

\section*{Data Availability}
Data is provided within the manuscript.

\section*{Funding}
This work was supported and funded by the DOE ARPA-E Program under the award number DE-AR0001578.

\section*{Declarations}
\vspace{5mm}

\section*{Competing interests}
The authors declare no competing interests

\section*{Additional information}
Correspondence and requests for materials should be addressed to G.T. or W.R.K.

\end{document}


\section*{Supplementary Information}

\setcounter{table}{0}
\renewcommand{\tablename}{Table}
\renewcommand{\thetable}{S\arabic{table}}

\setcounter{figure}{0}
\renewcommand{\figurename}{Figure}
\renewcommand{\thefigure}{S\arabic{figure}}

\begin{table}[ht]
    \caption{\label{tin}Initial isotopic composition of tin (6.9 g/cm$^3$) \cite{ref:wigeland}.}
    \centering
    \begin{tabular}{cccc}\toprule 
    Isotope&   Half-Life (days)& Mass Fraction
\\ \midrule
    Sn-115& stable& $0.4\%$ \\
    Sn-116& stable& $5.57\%$ \\
    Sn-117& stable& $9.55\%$ \\
    Sn-118& stable& $9.76\%$ \\
    Sn-119& stable& $9.81\%$ \\
    Sn-120& stable& $9.78\%$ \\
    Sn-122& stable& $10.69\%$ \\
    Sn-124& stable& $14.82\%$ \\
    Sn-126& 8.4E+07 & $29.57\%$ \\
    \bottomrule
    \end{tabular}
\end{table}

\begin{table}[ht]
    \caption{\label{zirconium}Initial isotopic composition of zirconium (6.5 g/cm$^3$) \cite{ref:wigeland}.}
    \centering
    \begin{tabular}{cccc}\toprule 
    Isotope&   Half-Life (days)& Mass Fraction
\\ \midrule
    Zr-90& stable& $1.73\%$ \\
    Zr-91& stable& $13.34\%$ \\
    Zr-92& stable& $16.22\%$ \\
    Zr-93& 5.59E+08& $20.26\%$ \\
    Zr-94& stable& $21.97\%$ \\
    Zr-96& 7.31E+21& $26.48\%$ \\
    \bottomrule
    \end{tabular}
\end{table}

\begin{table}[ht]
    \caption{\label{selenium}Initial isotopic composition of selenium (4.3 g/cm$^3$) \cite{ref:wigeland}.}
    \centering
    \begin{tabular}{cccc}\toprule 
    Isotope&   Half-Life (days)& Mass Fraction
\\ \midrule
    Se-77& stable& $3.44\%$ \\
    Se-78& stable& $6.53\%$ \\
    Se-79& 1.08E+08& $13.88\%$ \\
    Se-80& stable& $21.31\%$ \\
    Se-82& stable& $54.79\%$ \\
    \bottomrule
    \end{tabular}
\end{table}

\begin{table}[ht]
    \caption{\label{cesium}Initial isotopic composition of cesium (4.5 g/cm$^3$) \cite{ref:wigeland}.}
    \centering
    \begin{tabular}{cccc}\toprule 
    Isotope&   Half-Life (days)& Mass Fraction
\\ \midrule
    Cs-133& stable& $32.87\%$ \\
    Cs-134& 754.32& $0.33\%$ \\
    Cs-135& 8.4E+08& $38.06\%$ \\
    Cs-137& 1.10E+04& $28.74\%$ \\
    \bottomrule
    \end{tabular}
\end{table}

\begin{table}[ht]
    \caption{\label{iodine}Initial isotopic composition of iodine (4.5 g/cm$^3$) \cite{ref:wigeland}.}
    \centering
    \begin{tabular}{cccc}\toprule 
    Isotope&   Half-Life (days)& Mass Fraction
\\ \midrule
    I-127& stable& $27.85\%$ \\
    I-129& 5.73E+09 & $72.15\%$ \\
    \bottomrule
    \end{tabular}
\end{table}

\begin{figure}
    \centering
    \includegraphics[width=1.0\linewidth]{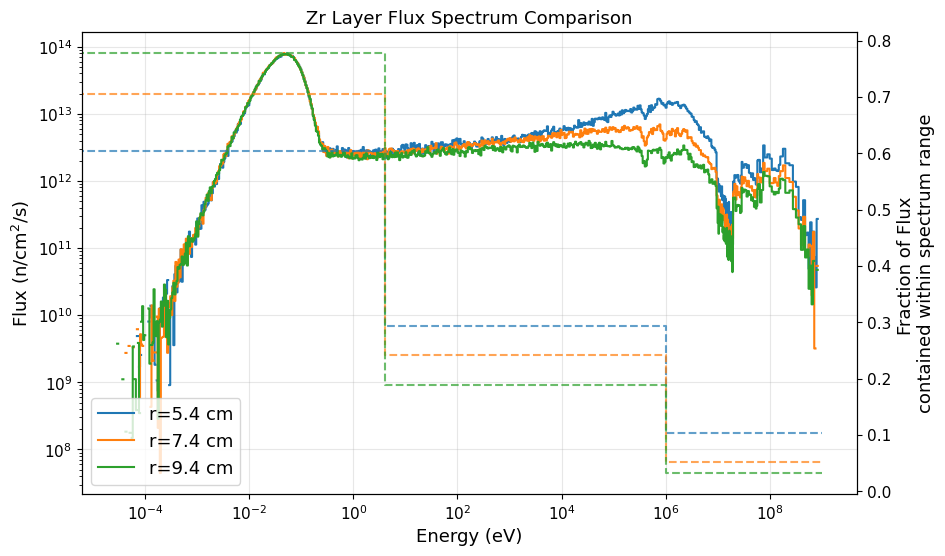}
    \caption{708-group neutron flux spectrum in a pure D$_2$O system at different radii. The left axis is flux magnitude, and the right axis is the fraction of the total flux that falls within each energy range, the ranges being 0-4 eV (thermal), 4-1e6 eV (epithermal), and 1e6-1e9 eV (fast). All fluxes were tallied over equal-volume regions. These radii correspond to layers 0, 5, and 10 in Figure 2 of the main text.}
    \label{fig:water-flux}
\end{figure}

\begin{figure}
    \centering
    \includegraphics[width=1.0\linewidth]{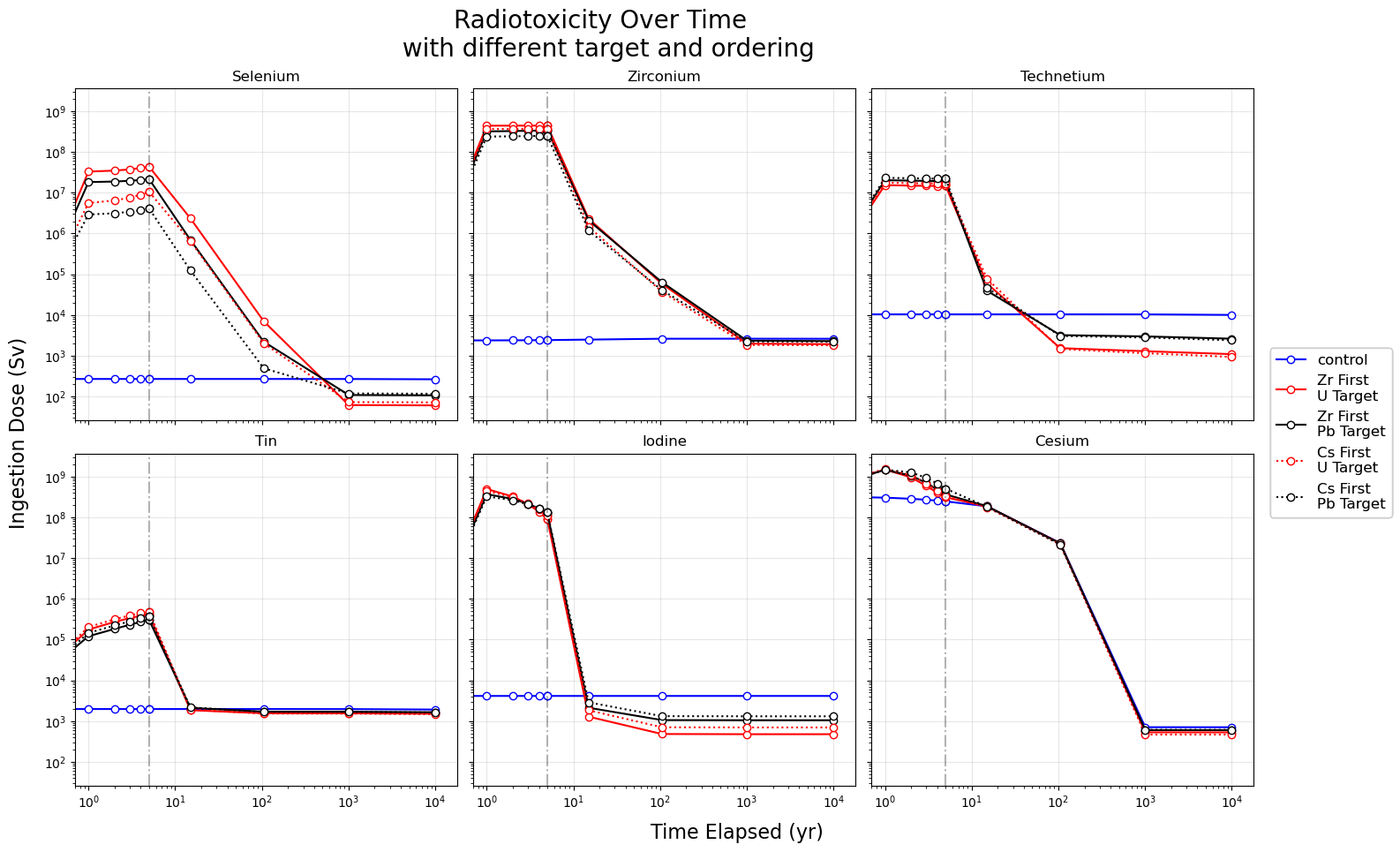}
    \caption{Radiotoxicity over time for each material in the combined geometry with different target material and ordering. The vertical line delineates when neutron transmutation ends. The control value is the radiotoxicity of the material over time under no neutron flux.}
    \label{fig:rt-time}
\end{figure}

\begin{table}[!ht]
\centering
\begin{tabular}{c |c | c c c}
\hline
    \multirow{2}{*}{Target} & \multirow{2}{*}{Ordering} & \multirow{2}{*}{Element} &  Specific Ingestion Dose & Relative Radiotoxicity \\
   & & & (Sv/g) & compared to control \\ \hline
    \multirow{12}{*}{Pb} & \multirow{6}{*}{Zr 1st} & Se & 1.755E-2 & 0.404 \\
    & & Zr & 2.046E-2 & 0.894 \\
    & & Tc & 1.163E-1 & 0.288 \\
    & & Sn & 5.437E-1&0.860 \\
    & & I & 1.285E-1&0.254 \\
    & & Cs & 2.752E-2&0.849\\ \cline{2-5}
    & \multirow{6}{*}{Cs 1st} & Se & 1.916E-2&0.441  \\
    & & Zr & 1.956E-2&0.855 \\
    & & Tc & 1.090E-1&0.270 \\
    & & Sn & 5.376E-1&0.851  \\
    & & I & 1.602E-1 &0.317 \\
    & & Cs & 2.798E-2& 0.863 \\ \hline
    \multirow{12}{*}{U} & \multirow{6}{*}{Zr 1st} & Se & 9.985E-3&0.229  \\
    & & Zr & 1.732E-2&0.757 \\
    & & Tc & 5.042E-2&0.125  \\
    & & Sn & 4.982E-1&0.788  \\
    & & I & 5.833E-2&0.115  \\
    & & Cs & 2.427E-2&0.748  \\ \cline{2-5}
    & \multirow{6}{*}{Cs 1st} & Se & 1.176E-2& 0.271 \\
    & & Zr & 1.609E-2&0.703  \\
    & & Tc & 4.506E-2&0.112  \\
    & & Sn & 4.892E-1&0.774  \\
    & & I &8.543E-2 &0.169  \\
    & & Cs & 2.139E-2& 0.659 \\
    \hline
\end{tabular}
\caption{Magnitude of radiotoxicity based on ingestion for each element, target material, and ordering. Values are after 5 years of irradiation and 1000 years of cooling. Relative value is in reference to the radiotoxicity of the material with no irradiation.}
\label{tab:rt-difference}
\end{table}

\begin{figure}
    \centering
    \includegraphics[width=\linewidth]{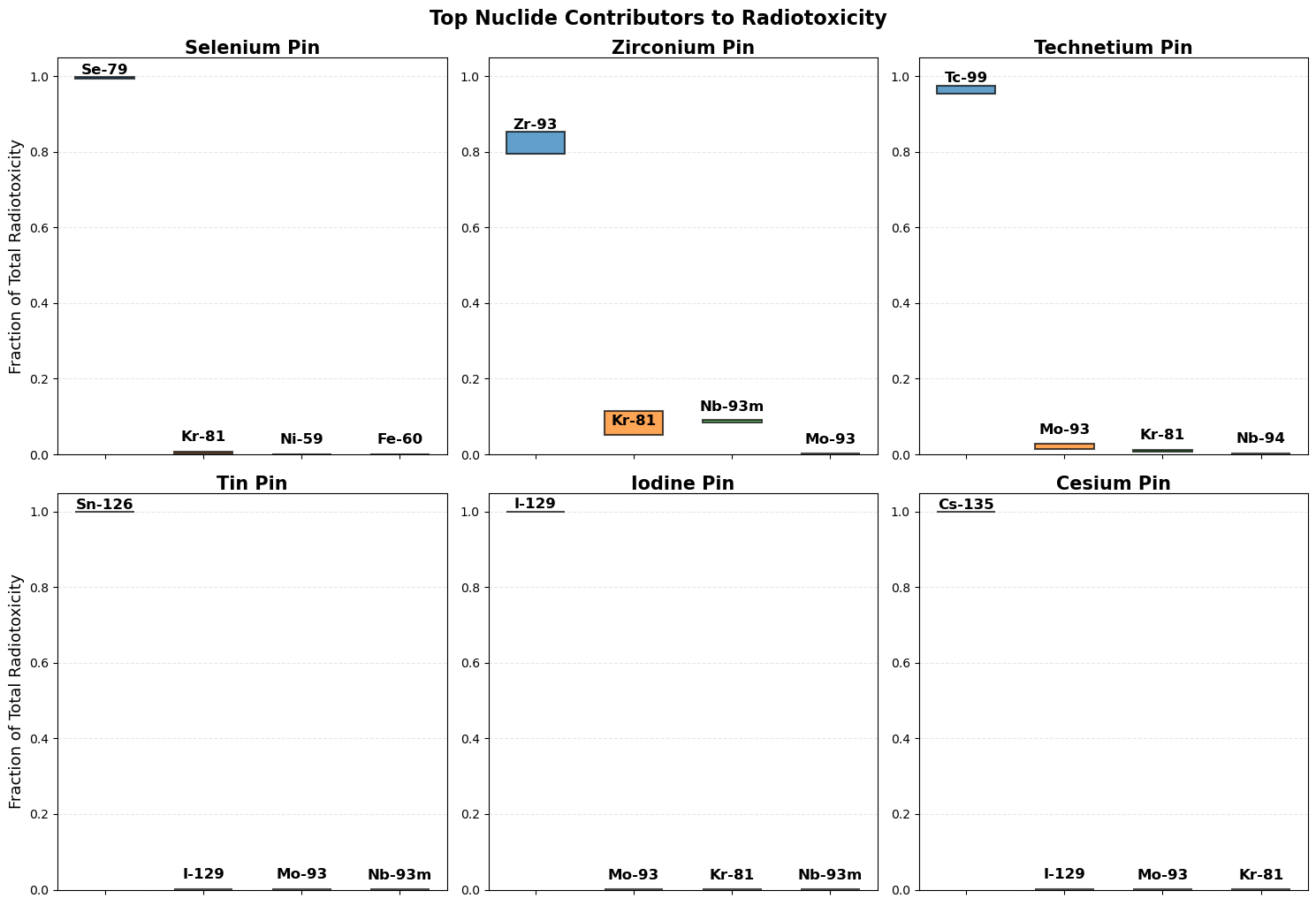}
    \caption{Fraction of the total radiotoxicity due to the top four contributing nuclides in the pin material. The width of each dataset represents the span of the values from different target material and element ordering. All values are from 5 years of irradiation and 1000 years of cooling.}
    \label{fig:rt-nucs}
\end{figure}

\newpage
\bibliography{main}